\documentclass[amsmath,amssymb,proc,secnumarabic,preprint]{revtex4}

\usepackage{graphicx}
\begin{document}

\title{Dynamics of graphene growth on a metal surface: a time-dependent photoemission study}

\author{Alexander~Gr\"{u}neis$^{1,2}$, Kurt Kummer$^{3}$ and Denis V. Vyalikh$^{3}$}

\affiliation{$^{1}$Faculty of Physics, University of Vienna,
Boltzmanngasse 5, A-1090 Vienna,
Austria}\affiliation{$^{2}$IFW--Dresden, P.O. Box 270116, D-01171
Dresden, Germany}\affiliation{$^{3}$Institute of Solid State Physics, Dresden University of Technology, D-01062 Dresden, Germany}


\begin{abstract}
Applying time-dependent photoemission we unravel the graphene
growth process on a metallic surface by chemical vapor deposition
(CVD). Graphene CVD growth is in stark contrast to the standard
growth process of two--dimensional films because it is
self-limiting and stops as soon as a monolayer graphene has been
synthesized. Most importantly, a novel phase of metastable
graphene was discovered that is characterized by permanent and
simultaneous construction and deconstruction. The high quality and
large area graphene flakes are characterized by angle-resolved
photoemission proofing that they are indeed monolayer and cover
the whole 1$\times$1~cm Nickel substrate. These findings are of
high relevance to the intensive search for reliable synthesis
methods for large graphene flakes of controlled layer number.
\end{abstract}
\maketitle

Its astonishing electronic properties have placed graphene, a planar sheet of carbon atoms packed in honeycomb structure, in the focus of considerable current research efforts.\cite{novoselov04-graphite,geim07-review} The occurrence of Dirac Fermions at low energies and the remarkably high electron mobility have raised high expectations regarding its future use as an active element in nanoelectronics and hybrid materials. Moreover, graphene is highly promising to play a crucial role in future spintronic applications and can serve as an effective oxidation protection when grown epitaxially on metal surfaces.\cite{Dedkov08-apl1}

Naturally these findings have stimulated the development of
efficient and reliable synthesis protocols for high-quality
graphene layers. To date several techniques were established: (i)
precipitation from silicon carbide~\cite{seyller06-sic}, (ii)
mechanical exfoliation from graphite~\cite{dai08-exfoliation},
(iii) reduction of exfoliated graphite oxide~\cite{Niyogi06JOTACS,
Gilje07NL, Stankovich06N}, (iv) thermal expansion of graphite
oxide~\cite{Schniepp06JPCB}, (v) laser
desorption~\cite{rader06-nm}, and (vi) growth by chemical vapor
deposition (CVD) on metal surfaces~\cite{alex07-graphenenickel,
nagashima94-graphene, oshima97-grapheneposition}. The latter is
especially interesting for ferromagnetic materials and it was
demonstrated that spin polarization of a Ni(111) substrate can be
cloned almost completely into a graphene
overlayer~\cite{karpan08-spinfilter}. This allows to design
sources of spin polarized electrons which are fully prevented from
aging when exposed to reactive gases. Despite these promising perspectives, little is known about the actual growth mechanism of graphene on metal surfaces~\cite{loginova08-njp}.

\begin{figure}
\includegraphics[width=15cm]{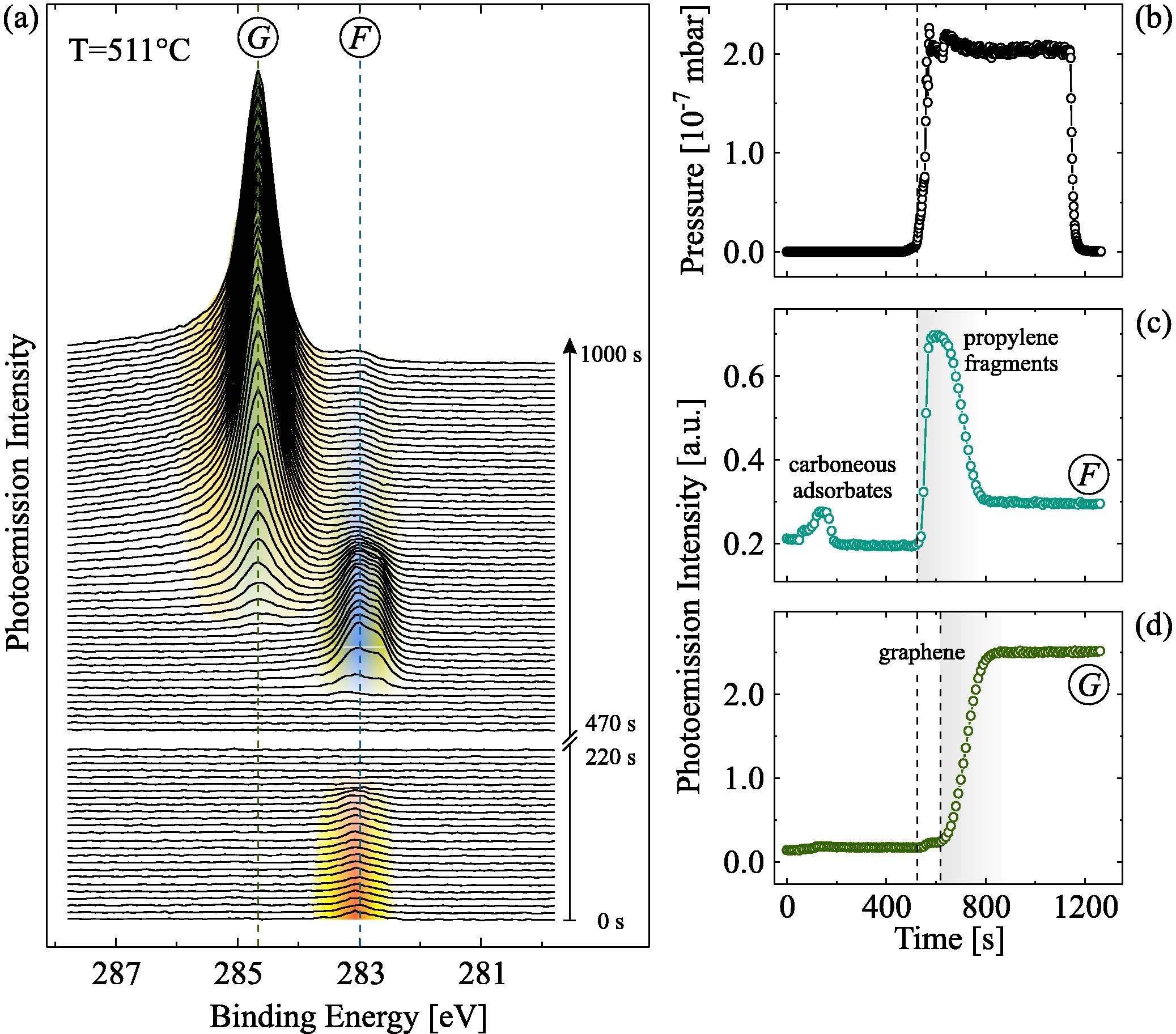}
\caption{(a) Time evolution of the PE intensity in the C~1\textit{s} region during the graphene growth. F and G mark the signals from C$_3$H$_6$ fragments and graphene. (b) Partial C$_3$H$_6$ pressure. (c) and (d) Intensity of the fragment and the graphene C~1$s$ PE signal integrated over $\pm$0.5~eV around their peak maximum.}
\label{fig:fig1}
\end{figure}

Here we show that the high spatial compatibility of the Ni(111) surface and the graphene sheet makes it a perfect system to study functional synthesis of monolayer graphene on metal substrates by a self limiting CVD process. The high crystallinity and large area growth of monolayer graphene is unambigously proven by angle--resolved photoemission spectroscopy (ARPES) of the electronic
band structure.  For CVD of the graphene layer, a freshly prepared Ni(111) substrate was heated and stabilized at the desired synthesis temperature first. Then propylene gas (C$_3$H$_6$), which served as the carbon source, was introduced into the chamber. Doing so the gas pressure was adjusted to $\rm 2\times 10^{-7}$~mbar using a leak valve. In the incipient reaction of propylene with the Ni surface a graphene monolayer is formed as evidenced by the appearance of a single $\pi$ band by ARPES as will be
shown later.
During the whole synthesis procedure the C~1\textit{s} signal was recorded together with the C$_3$H$_6$ gas pressure and the substrate temperature. This complete set of information allowed us to monitor the full growth process dependent on time.
One time-dependent data set during the graphene CVD is shown
in~\ref{fig:fig1}. Starting from a blank Ni film kept at room
temperature it shows the time evolution of (a) the raw PE signal
in the C~1\textit{s} range, (b) the C$_3$H$_6$ partial pressure,
(c) and (d) the C~1$s$ signal intensity integrated for a range of
$\pm$0.5~eV around the blue and green lines in (a), respectively.
Considering the C~1\textit{s} spectra it is evident that in the beginning the
prepared Ni(111) surface still shows small signs of carboneous
contaminations at 283~eV binding energy (BE). They however are
fully removed during the first 200~s when the substrate
temperature is raised to the synthesis temperature for the
graphene growth. After the substrate is stabilized in
temperature the leak valve to the C$_3$H$_6$ gas inlet is
opened (this point corresponds to a time $t\sim$500s). Immediately with
increasing C$_3$H$_6$ pressure a distinct peak at 283~eV BE
arises. Only at $t\sim$600~s the graphene related peak at 284.7~eV
starts to grow (see the dashed lines in ~\ref{fig:fig1}b--d).

Three origins of the 283~eV peak are conceivable: (i) unfragmented
C$_3$H$_6$, (ii) surface nickel carbide and (iii) C$_3$H$_6$
fragments. The former two, however, are not supported by
additionally performed experiments. First, producing a C$_3$H$_6$
film on Ni(111) by room temperature adsorption we found the
C~1\textit{s} PE peak at $\sim$284~eV BE and not at 283~eV BE (not
shown). Secondly, we have deliberately synthesized nickel carbide
and proofed this by LEED as described previously
\cite{tanaka92-cl}. We indeed detected the C~1$s$ peak at 283~eV BE,
in agreement with literature values \cite{kovacs08-tsf} for nickel
carbide. But we were not able to convert the surface nickel
carbide film into graphene by heating. Instead we observed that
the nickel carbide film is stable and the peak at 283~eV did not
vanish nor did appear a graphene peak. Thus we conclude that the
observed peak at 283~eV BE is due to C$_3$H$_6$ fragments and in
agreement to previous literatures that report a very similar
experiment of carbon monoxide decomposition on Rh, we assign it partly to
atomic carbon~\cite{mongeot06-prl}.  Another contribution to this peak
might be edge atoms of the growing graphene film. Indeed, this would explain why
the intensity decreases as the graphene layer forms and has fewer edge atoms. 
Our assignment is also consistent with the fact that we did not observe any sign of C-H
bonds, which appear as a sideband in the C~1$s$ peak at $\sim$1eV
higher binding energy than the graphene peak~\cite{nikitin05-h}.

Looking at \ref{fig:fig1}c and d one recognizes that during construction of the graphene network a major part of the fragments vanish, i.e. are transformed into graphene. When the graphene growth is completed only a minor fraction of fragments remains. For all recorded spectra the beginning of both the graphene peak rise and the fragment peak decline coincide in time. Comparing the maximum intensity of the fragment peak ($\sim$0.7~a.u., \ref{fig:fig1}c) and graphene peak ($\sim$2.5~a.u., \ref{fig:fig1}d) it becomes apparent that there must be an additional major path of carbon incorporation to achieve a complete monolayer. We believe that graphene nucleates initially from fragments that sit on edges and defects of the Ni surface. The part missing to a full graphene monolayer grows by attachment of carbons.

\begin{figure}
\includegraphics[width=15cm]{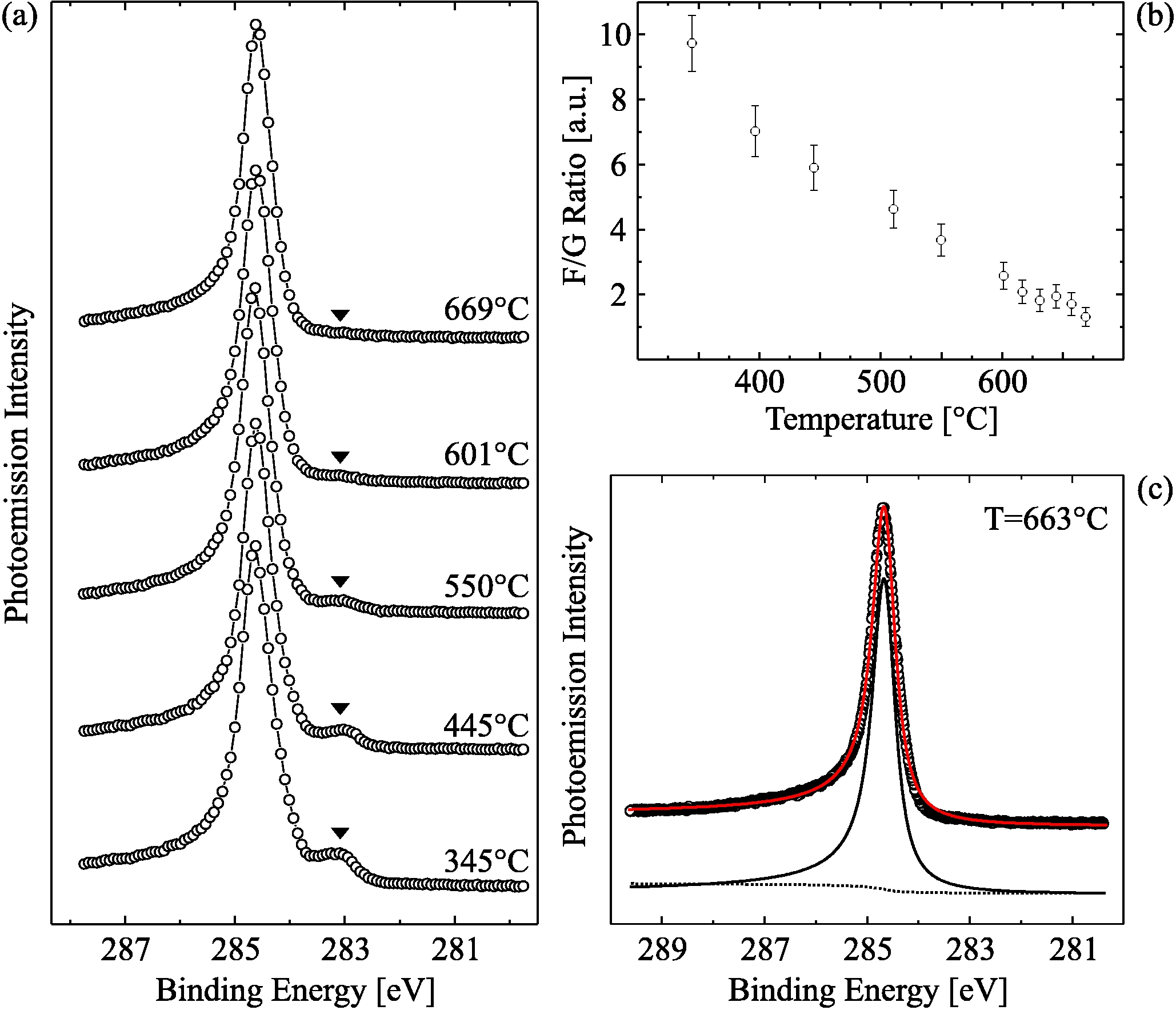}
\caption{(a) C~1\textit{s} PE spectrum of the fully grown graphene layer for different temperatures. (b) Fragment to graphene PE intensity ratio as a function of synthesis temperature. (c) High resolution C~1\textit{s} spectrum for graphene on Ni(111) along with a Doniach-Sunjic lineshape analysis (red line).}
\label{fig:tempdep}
\end{figure}

The percentage of fragments in the final graphene layer depends
considerably on the synthesis temperature as shown in
\ref{fig:tempdep}a. While clearly present for lower synthesis
temperatures a distinct fragment peak is not observed beyond
$\sim$600$^\circ$C. This highlights that high temperatures result
in an efficient conversion of fragments to graphene. The
quantitative comparison of the fragment PE and the graphene PE
intensities in \ref{fig:tempdep}b reveals a decline in their ratio
from 10\% to about 1\% when going from 345$^\circ$C to
669$^\circ$C. The high structural quality of the graphene layer
synthesized at 669$^\circ$C is confirmed by least-squares fit
analysis of the C~1\textit{s} PE spectrum. Using a single
component with Doniach-Sunjic lineshape we obtained
$\Gamma_l$=216~meV and $\alpha=0.1$ for the intrinsic line width
and asymmetry parameter, respectively which is in agreement to
previously reported values.\cite{preobrajenski08-prb}

\begin{figure}
\includegraphics[width=15cm]{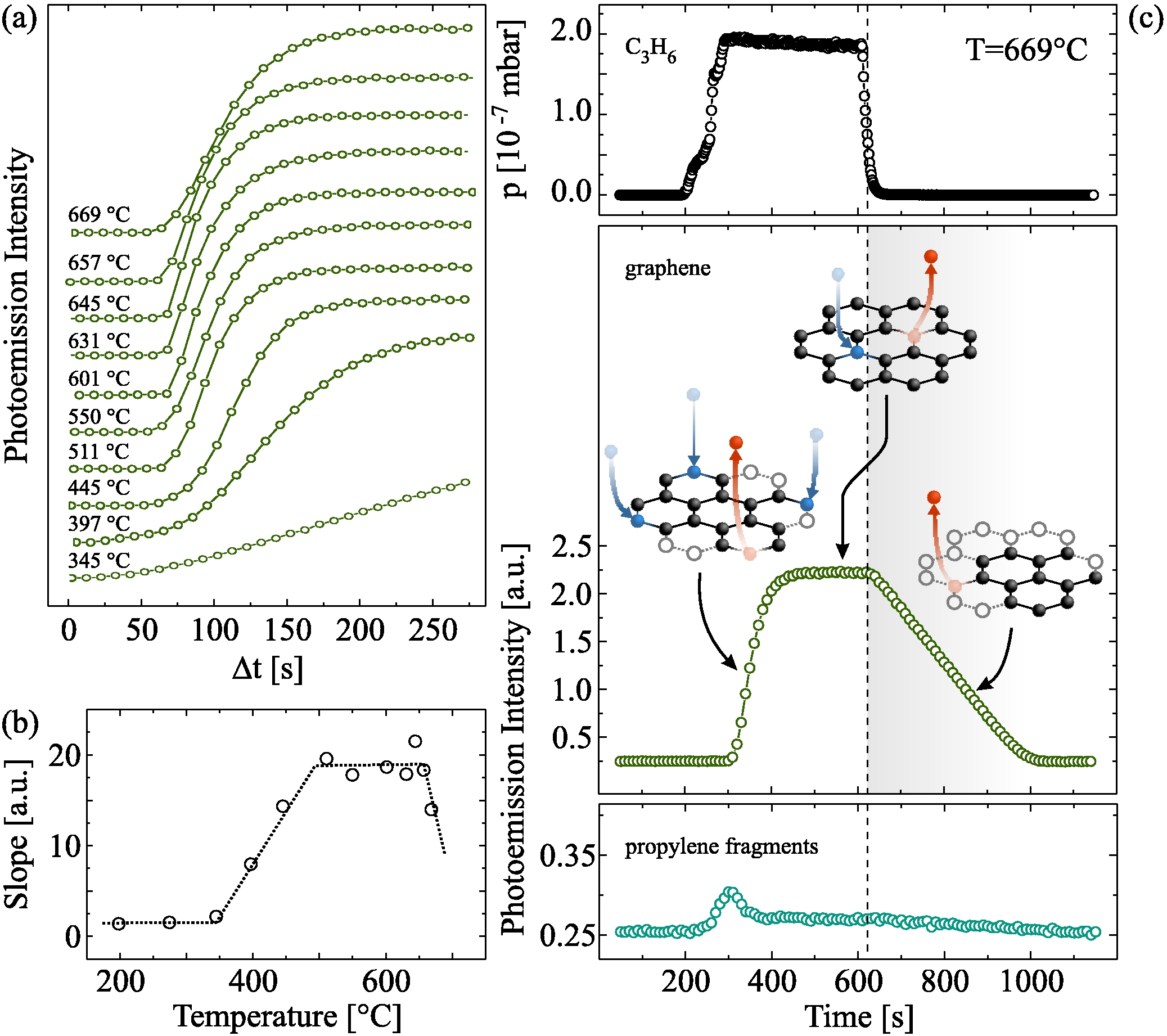}
\caption{(a) Time evolution of the graphene C~1\textit{s} intensity for different temperatures. (b) Graphene growth rate as a function of synthesis temperature. (c) Graphene C~1\textit{s} intensity spectrum for T=669$^\circ$C indicating three consecutive development stages: rapid growth, a metastable phase, and desorption or diffusion.}
\label{fig:metastable}
\end{figure}

The synthesis temperature is decisive not only for the structural
quality of the graphene layer, it furthermore governs the graphene
growth rate. In \ref{fig:metastable}(a) the time evolution of the
graphene PE intensity is shown for increasing temperatures. For
the purpose of comparison the curves were aligned to each another
on the time scale. Taking the slope of the C~1$s$ intensity of
graphene at the turning point of the curve as a measure for the
growth rate we found that the latter has a synthesis temperature
dependence as shown in \ref{fig:metastable}b. For temperatures
below 350$^\circ$C no extensive graphene growth occurs. Then a
rapid increase in the growth rate is observed when the synthesis
temperature is raised to about 500$^\circ$C. Beginning from
500$^\circ$C the growth rate remains constant before it appears to
decline again above 650$^\circ$C. In contrast to graphene layers
synthesized at lower temperatures, we found that those constructed
at 657$^\circ$C and 669$^\circ$C exist only in a metastable phase.

In \ref{fig:metastable}c we show a full data set of propylene
pressure and fragment and graphene C~1$s$ PE intensities for
T=669$^\circ$C. When propylene is introduced into the chamber the
graphene grows up to completion of one monolayer. However, after a
closed graphene layer is reached and the gas inlet closed, the
graphene C~1$s$ intensity decreases, unless the temperature has
been lowered before. The graphene layer completely disappeares in
400s after the carbon supply has been turned off.

Two possible reasons for the disappearance of graphene carbon
atoms are conceivable. First, it may be possible that at such high
temperatures carbon atoms in graphene react with residual gases in
the chamber, e.g. hydrogen, and form hydrocarbons which can easily
desorb. Secondly, diffusion of carbon atoms into the bulk Ni could
set in for the highest applied temperatures \cite{siegel03-prb,
lander52-jap, diamond67-tms}. The question, whether desorption or
diffusion is dominating remains subject to further investigations.
We wish to point out that after high-temperature deconstruction of
the graphene layer no notable amount of carbon remains on the
surface. In fact, after repeated synthesis and heating procedures
we always found a pristine Ni surface, ready for a new cycle of
graphene growth.

\begin{figure}
\includegraphics[width=15cm]{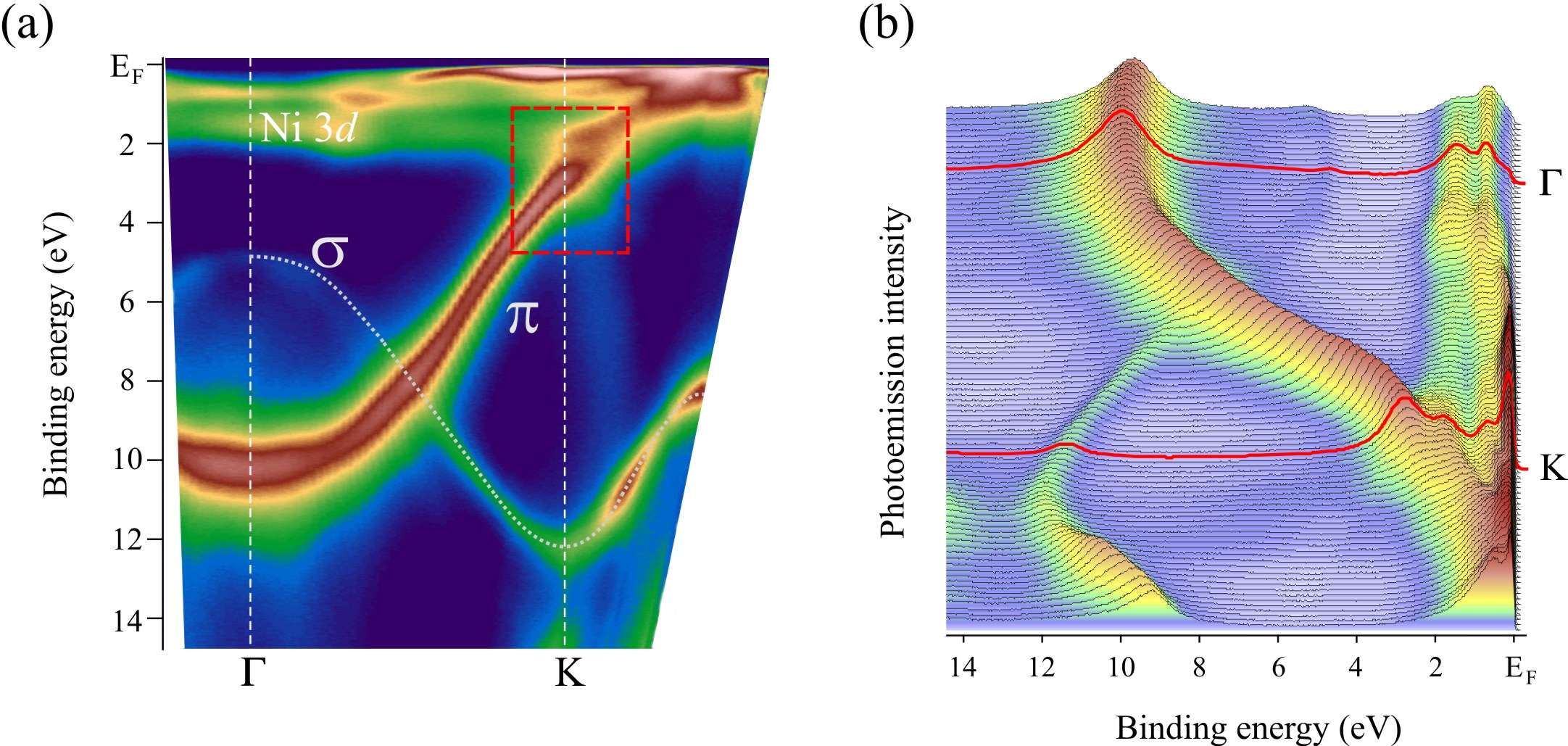}
\caption{(a) Band structure mapping by ARPES of a graphene monolayer synthesized at 550~$\rm ^o$C. The graphene dervived $\pi$ and
$\sigma$ bands and the Ni~$3d$ bands are depicted. The region around $K$ denoted by a dashed rectangle exhibits the gap in
the $\pi$ band structure, which is due to substrate interaction. (b) The raw ARPES spectra: red lines denote scans taken at $\Gamma$ and $K$ points, respectively.}
\label{fig:fig4}
\end{figure}

Finally, we demonstrate the high crystallinity of a graphene monolayer synthesized at 550$\rm ^o$C
by performing a mapping of the electronic band structure using ARPES.
The measured band structure between $\Gamma$ and $K$ points
in the two--dimensional Brillouin zone is shown in \ref{fig:fig4}a.
The appearance of $\pi$ and $\sigma$ bands is a clear indication for long range crystallinity
and few defects in the honeycomb lattice. Furthermore a gap at $K$ point appears which is a result
of substrate interaction and hybridization of C~$2p_z$ and Ni~$3d_{3z^2-r^2}$ orbitals as disscussed previously~\cite{alex07-graphenenickel}.
In \ref{fig:fig4}b the corresponding raw spectra are depicted. It can be seen from the ARPES results
that only one $\pi$ valence band is visible. This demonstrates unambigously that
we have indeed synthesized monolayer graphene as a graphene bilayer would have two $\pi$ bands. The ARPES spectra are also strongly supporting the observation from the time--resolved photoemission that the catalytic activity diminishes after the growth of one monolayer.

In summary, applying time-resolved PES we were able to observe and characterize the dependence of graphene quality and growth rate on the synthesis temperature, for the first time. For high temperatures above 650$^\circ$C we found the graphene layer in a metastable state characterized by permanent simultaneous construction and deconstruction. The high crystallinity of the
synthesized graphene monolayers was confirmed by a mapping of the electronic band structure by ARPES.

\section*{Experimental}
The Ni films were grown onto a W(110) single crystal which could be heated from the backside by electron bombardment. The temperature was monitored using a thermocouple which was spot welded on the back side of the tungsten crystal to ensure good thermal contact. During deposition the tungsten crystal was heated to constant 150$^\circ$C from the backside by electron beam. The W(110) surface was cleaned by repeated cycles of short flashes up to 1700$^\circ$C and annealing in oxygen at 1000$^\circ$C. Subsequently, a Ni film of 10~nm thickness was deposited on the W(110) surface by electron beam evaporation from a Ni rod (99.9\% purity). The Ni film thickness was monitored by a quartz microbalance. Nickel grows epitaxially in (111) fashion on W(110)~\cite{gunterodt88-ni111w110} and we further employ it as a catalytic template for graphene CVD due to its small lattice mismatch. Low energy electron diffraction (LEED) and photoelectron spectroscopy (PES) were utilized to check the W(110) and Ni(111) surfaces for order and cleanliness~\cite{alex07-graphenenickel, gunterodt88-ni111w110}. For all experiments the base pressure before dosing C$_3$H$_6$ was better than $5\times 10^{-10}$~mbar, during Ni evaporation better than $2\times 10^{-9}$~mbar.
Time-dependent experiments were carried out at the SuperESCA beamline of the ELETTRA synchrotron (Trieste, Italy).
The C~1\textit{s} spectra were recorded at h$\nu$=400~eV photon energy. The total experimental resolution was $\sim$~180 meV and the spot size on the sample was 200$\times$30$\mu$m.
The electronic band structure was measured with ARPES at the IFW-Dresden using a photoemission spectrometer
equipped with a Scienta SES-200 hemispherical electron-energy analyzer and a high-flux He-resonance lamp (Gammadata VUV-5010). All
ARPES spectra were acquired at a photon energy of h$\nu$=40.8 eV (He II$\alpha$) with an angular resolution of 0.3$^{\circ}$ and
a total-system energy resolution of 50 meV.


\section*{Acknowledgements} A.Gr\"uneis acknowledges an APART fellowship from the Austrian Academy of
Sciences and a Marie Curie Individual
Fellowship ("COMTRANS") and travel support from the EU. D.V.
acknowledges the Deutsche Forschungsgemeinschaft (Grant VY 64/1-1). We are grateful to A.~Goldoni, S.~Lizzit and P. Vilmercati for their support at the SuperESCA beamline and to R.~H{\"u}bel for his support at the IFW-Dresden.

\clearpage

\clearpage

\clearpage

\clearpage


\clearpage
\begin{thebibliography}{1}

\bibitem{novoselov04-graphite}
Novoselov, K. S.; Geim, A.K.; Morozov, S.V.; Jiang, D.; Zhang, Y.; Dubonos, S.V.; Grigorieva, I.V.; Firsov, A.A. \textsl{Science} \textbf{2004}, \textsl{306}, 666.

\bibitem{geim07-review}
Geim, A.K.; Novoselov, K.S. \textsl{Nature Mat.} \textbf{2007}, \textsl{6}, 183.

\bibitem{Dedkov08-apl1}
Dedkov,~Y.; Fonin,~M.; R\"udiger,~U.; Laubschat,~C. \textsl{Appl. Phys. Lett.}
  \textbf{2008}, \textsl{93}, 022509.

\bibitem{seyller06-sic}
Seyller,~T.; Emtsev,~K.; Gao,~K.; Speck,~F.; Ley,~L.; Tadich,~A.; Broekman,~L.;
  Riley,~J.; Leckey,~R.; Rader,~O.; Varykhalov,~A.; Shikin,~A. \textsl{Surf.
  Sci.} \textbf{2006}, \textsl{600}, 3906.

\bibitem{dai08-exfoliation}
Xiaolin,~L.; Guangyu,~Z.; Xuedong,~B.; Xiaoming,~S.; Xinran,~W.; Enge,~W.;
  Dai,~H. \textsl{Nature Nanotech.} \textbf{2008}, \textsl{3}, 538.

\bibitem{Niyogi06JOTACS}
Niyogi,~S.; Bekyarova,~E.; Itkis,~M.~E.; McWilliams,~J.~L.; Hamon,~M.~A.;
  Haddon,~R.~C. \textsl{J. Am. Chem. Soc.} \textbf{2006},
  \textsl{128}, 7720.

\bibitem{Gilje07NL}
Gilje,~S.; Han,~S.; Wang,~M.; Wang,~K.~L.; Kaner,~R.~B. \textsl{Nano Letters}
  \textbf{2007}, \textsl{7}, 3394.

\bibitem{Stankovich06N}
Stankovich,~S.; Dikin,~D.~A.; Dommett,~G. H.~B.; Kohlhaas,~K.~M.;
  Zimney,~E.~J.; Stach,~E.~A.; Piner,~R.~D.; Nguyen,~S.~T.; Ruoff,~R.~S.
  \textsl{Nature} \textbf{2006}, \textsl{442}, 282.

\bibitem{Schniepp06JPCB}
Schniepp,~H.; Li,~J.-L.; McAllister,~M.; Sai,~H.; Herrera-Alonso,~M.;
  Adamson,~D.; Prud'homme,~R.K.; Car,~R.; Saville,~D.; Aksay,~I. \textsl{J. Phys.
  Chem. B} \textbf{2006}, \textsl{110}, 8535.

\bibitem{rader06-nm}
Rader, H.; Rouhanipour, A.; Talarico, A.; Palermo, V.; Samori, O.; Mullen, K. \textsl{Nature Mat.} \textbf{2006}, \textit{5}, 276.

\bibitem{alex07-graphenenickel}
Gr\"uneis,~A.; Vyalikh,~D. \textsl{Phys. Rev. B} \textbf{2008}, \textsl{77}, 193401.

\bibitem{nagashima94-graphene}
Nagashima, A.; Tejima, N.; Oshima, C. \textsl{Phys. Rev. B} \textbf{1994}, \textsl{50}, 17487.

\bibitem{oshima97-grapheneposition}
Gamo, Y.; Nagashima, A.; Wakabayashi, M.; Oshima, C. \textsl{Surf. Sci.} \textbf{1997}, \textsl{374}, 61.

\bibitem{karpan08-spinfilter}
Karpan, V.M.; Giovannetti, G.; Khomyakov, P.A.; Talanana, M.; Starikov, A.A.; Zwierzycki, M.; van den Brink, J.; Brocks, G.; Kelly, P.J. \textsl{Phys. Rev. Lett.} \textbf{2007}, \textsl{99}, 176602.

\bibitem{loginova08-njp}
Loginova, E.; Bartelt, N.C.; Feibelman, P.J.; McCarty, K.F. \textsl{New J. Phys.} \textbf{2008}, \textsl{10}, 093026.

\bibitem{tanaka92-cl}
Tanaka, K.; Hirano, H. \textsl{Catalysis Lett.} \textbf{1992}, \textsl{12}, 1.

\bibitem{kovacs08-tsf}
Kov\'acs, Gy.J.; Bert\'oti, I.; Radn\'oczi, G. \textsl{Thin Solid Films} \textbf{2008}, \textsl{516}, 7942.

\bibitem{mongeot06-prl}
Mongeot, F.; Toma, A.; Molle, A.; Lizzit, S.; Petaccia, L.; Baraldi, A. \textsl{Phys. Rev. Lett.} \textbf{2006}, \textsl{97}, 056103.

\bibitem{nikitin05-h}
Nikitin,~A.; Ogasawara,~H.; Mann,~D.; Denecke,~R.; Zhang,~Z.; Dai,~H.; Cho,~K.;
  Nilsson,~A. \textsl{Phys. Rev. Lett.} \textbf{2005}, \textsl{95}, 225507.

\bibitem{preobrajenski08-prb}
Preobrajenski, A.B.; Ng, M.L., Vinogradov, A.S., Martensson, N. \textsl{Phys. Rev. B} \textbf{2008}, \textsl{78}, 073401.

\bibitem{siegel03-prb}
Siegel, D.J.; Hamilton, J.C. \textsl{Phys. Rev. B} \textbf{2003}, \textsl{68}, 094105.

\bibitem{lander52-jap}
Lander, J.; Kern, H.; Beach, A. \textsl{J. Appl. Phys.} \textbf{1952}, \textsl{23}, 1305.

\bibitem{diamond67-tms}
Diamond, S.; Wert, C. \textsl{Trans. Metall. Soc. AIME} \textbf{1967}, \textsl{239}, 705.

\bibitem{gunterodt88-ni111w110}
K\"amper, K.-P.; Schmidt, W.; G\"untherodt, G.; Kuhlenbeck, H. \textsl{Phys. Rev. B} \textbf{1988}, \textsl{38}, 9451.

\end{thebibliography}
\end{document}